\documentclass[10pt,nofootinbib,twocolumn,aps,prl,letterpaper,showpacs,prd,longbibliography]{revtex4-1}

\usepackage{amsmath}
\usepackage{amsfonts}
\usepackage{amssymb}
\usepackage{bbm}
\usepackage[dvips]{graphicx}
\usepackage[dvips,bookmarks=false]{hyperref} 
\usepackage[applemac]{inputenc}

\hypersetup{pdfstartview=FitH,pdfhighlight=/O,colorlinks=false}

\begin{document}                

\newcommand{\mc}[1]{\mathcal{#1}}
\newcommand{\mbb}[1]{\mathbbm{#1}}
\newcommand{\Tr}{{\rm Tr}}
\newcommand{\ket}[1]{|#1\rangle}
\newcommand{\varket}[1]{[#1\rangle}
\newcommand{\bra}[1]{\langle#1|}

\title{\large\bf Curvature in spinfoams}

\author{Elena Magliaro}
\email{magliaro@gravity.psu.edu}
\author{Claudio Perini}
\email{perini@gravity.psu.edu}
\affiliation{Institute for Gravitation and the Cosmos, Physics Department, Penn State, University Park, PA 16802-6300, USA}

\date{\small\today}

\begin{abstract}
We consider spinfoam quantum gravity. We show in a simple case that the amplitude projects over a nontrivial (curved) classical geometry. This suggests that, at least for spinfoams without bubbles and for large values of the boundary spins, the amplitude takes the form of a path integral over Regge metrics, thus enforcing discrete Einstein equations in the classical limit. The result relies crucially on a new interpretation of the semiclassical limit for the amplitudes truncated to a fixed 2-complex.
\end{abstract}

\maketitle

\section{Introduction} Spinfoams \cite{Baez:1997zt,Reisenberger:2000fy,Perez:2003vx} are a tentative covariant quantization of general relativity. They provide transition amplitudes between quantum states of 3-geometry, in the form of a Misner-Hawking sum over virtual geometries \cite{Misner:1957wq,Hawking:1979zw}. In the so-called `new models' \cite{Engle:2007wy,Freidel:2007py}, intermediate quantum states are the ones of canonical loop quantum gravity, the $SU(2)$ spin-network states. The physical picture emerging from the spinfoam gravity is that of a discrete, combinatorial spacetime structure (a quantum foam of virtual geometries), where the Plank scale plays the role of a natural minimal length. Spinfoams are the result of the quantization of general relativity formulated as a constrained BF theory. In the BF theory with gauge group $G$, the basic variables are a 2-form $B$ and a connection 1-form $\omega$, both valued in the Lie algebra of $G$. The equations of motion of BF theory impose the flatness of the connection
\begin{align}
F(\omega)=d\omega+\omega\wedge\omega=0
\end{align}
where $F$ is the curvature. So this theory has no local degrees of freedom. The simplicity constraints on the field $B$ turn the topological BF theory into vacuum general relativity, described by the Holst action
\begin{align}\label{Sintro}
S=\int{}^*(e\wedge e)\wedge F(\omega)+\frac{1}{\gamma}\int (e\wedge e)\wedge F(\omega)
\end{align}
with gauge group $SO(1,3)$ (or $SO(4)$ in the Euclidean signature), where the second term vanishes by using the equations of motion. The first term is the Einstein-Hilbert-Palatini action for general relativity in terms of the cotetrad $e$ and the connection $\omega$, regarded as independent variables. The real number $\gamma\neq0$ is called Barbero-Immirzi parameter. Spinfoam models provide a Feynman path integral, or state sum, based on a discretization of \eqref{Sintro} over a 2-complex. The spinfoam theory is sufficiently simple \cite{Rovelli:2010bf} and possesses the correct symmetries \cite{Rovelli:2010ed}, and recently  has been successfully coupled to matter fields \cite{Bianchi:2010bn,Han:2011as}. Furthermore, the large distance analysis was able to extract the correct low-energy physics in some simple cases \cite{Bianchi:2006uf,Alesci:2007tx,Bianchi:2009ri,Bianchi:2010zs}. 

A major open problem is to show that the spinfoam Feynman path integral is able to reproduce Einstein equations when a semiclassical expansion is performed. What we would like to have is the spinfoam version of the following semiclassical expansion of the gravitational path integral
\begin{align}
\int Dg_{\mu\nu}e^{\frac{i}{\hbar}S_{EH}(g_{\mu\nu})}\sim e^{\frac{i}{\hbar}S_{EH}(g^0_{\mu\nu})}
\end{align}
in the classical limit $\hbar\rightarrow 0$, where $S_{EH}$ is the Eintein-Hilber action for general relativity and on the right hand side it is evaluated on the classical solution of the equations of motion determined by the boundary conditions specified on the metric field $g_{\mu\nu}$.

In particular, in the spinfoam framework it is not clear if besides flat geometries (vanishing of the Riemann tensor), also curved geometries are allowed (vanishing of the Ricci tensor) in the classical limit. It could be that the local degrees of freedom of general relativity are lost and the theory is classically equivalent to a topological theory. If this is not the case, it would be a further test of the consistency of the spinfoam framework.

In order to consider curved spacetimes, we have to go beyond the single vertex, or single 4-simplex approximation, where many calculations have been done.

In this paper, we introduce a simple triangulation of spacetime with three vertices and one internal face (the hinge, where curvature is concentrated) and show that the spinfoam boundary amplitude is peaked on geometries with nonzero curvature. We work in natural units $G=\hbar=c=1$.

\section{The spinfoam amplitude}
We consider the spinfoam amplitude \cite{Engle:2007wy,Freidel:2007py} for a 2-complex $\sigma$ with boundary graph $\Gamma=\partial\sigma$. We restrict for simplicity to the Euclidean signature and to Barbero-Immirzi parameter $0<\gamma<1$, where formulas get simpler. For each face $f$ of $\sigma$ there is an associated integer spin $j_f$; the boundary, or external, faces cut the boundary over a link $l$ of $\Gamma$, with associated spin $j_l$. Faces are oriented and bounded by a cycle of edges $e$. Each edge bounding a face has a source vertex $s(e)$ and a target vertex $t(e)$, where source/target is relative to the orientation of the face. To each edge associate $SU(2)$ elements $n_{ef}$ ($f$ runs over the faces meeting at the edge $e$) and two source/target 
\begin{align}
Spin(4)\simeq SU(2)\times SU(2)\sim SO(4)
\end{align}
gauge group variables $g_{e,s(e)}$, $g_{e,t(e)}$. The variables $n_{ef}$ can also be interpreted as unit vectors $\vec n_{ef}$ in $\mathbbm R^3$, up to a phase ambiguity, by saying that $n$ is a rotation that brings a reference direction to the direction of $\vec n$. External edges puncture the boundary at one node of the boundary graph and are labeled\footnote{$l$ runs over the links $l$ surrounding the node $n$ of the graph.} by $n_{nl}\in SU(2)$. The set formed by the boundary graph $\Gamma$, spins $j_l$ and unit vectors $n_{nl}$ constitutes the boundary data.

The spinfoam amplitude for the 2-complex in the Bloch coherent state basis \cite{Livine:2007vk} is a functional of the boundary data defined as
\begin{align}
W(j_l,n_{nl})=\sum_{\{j_f\}}\int dg_{ve}\int dn_{ef}\prod_f P_f.
\label{amplitudejn}
\end{align}
The sum is over the internal spins $j_f\in \mathbbm N$, and the integrals are over the $Spin(4)$ gauge variables and $SU(2)$ variables labeling the internal edges. For an internal face, the face amplitude $P_f$ is given by
\begin{align}\label{Pf}
P_f=\text{tr}\,\vec\Pi_{e\in f} P_{ef}
\end{align}
where $\vec\Pi$ denotes the ordered product (according to the cycle of edges) and
\begin{align}\label{Pef}
P_{ef}= g_{e,{s(e)}}Y\varket{j_f,n_{ef}}\bra{j_f,n_{ef}}Y^\dagger g^{-1}_{e,t(e)}.
\end{align}
Here $\ket{j,n}$ is the $SU(2)$ Bloch coherent state \cite{Bloch:1946zza} for angular momentum\footnote{We introduced the notation $\varket{j,n}$ for the standard antilinear map $\epsilon$ applied to $\ket{j,n}$. In the standard basis, it is given by the symbol ${}^{j}\epsilon_{mm'}=(-1)^{j+m}\delta_{m,-m'}$.} along the direction $\vec n$. The map $Y$ gives the embedding of the $SU(2)$ irreducible representation space with spin $j$ into $SU(2)\times SU(2)$ representation space with spins $(j^+,j^-)$, where $j^\pm=\frac{1\pm\gamma}{2}j$. In the canonical basis it is given by the Clebsh-Gordan coefficient
\begin{align}
\bra{j^+j^-;m^+m^-}Y\ket{j;m}=C(j^+j^-j;m^+m^-m).
\end{align}
Using the factorization properties of $SU(2)$ coherent states we can write
\begin{align}
P_{ef}=P^+_{ef}\otimes P^-_{ef}
\end{align}
and
\begin{align}
P^\pm_{ef}=g^\pm_{e,s(e)}\varket{n_{ef}}^{\otimes2j^\pm_f}\bra{n_{ef}}^{\otimes2j^\pm_f}(g^\pm_{e,t(e)})^{-1} 
\end{align}
where we have split the $SO(4)$ variables $g$ in selfdual and antiselfdual rotations $g^+,g^-\in SU(2)$. For an external face, the formula for $P_f$ is the same except that for the edges ending on the boundary we have only `half' of \eqref{Pef}, namely
\begin{align}
P_{ef}=\bra{j_f,n_{ef}}g^{-1}_{e,t(e)}
\end{align}
or
\begin{align}
P_{ef}=g_{e,{s(e)}}Y\varket{j_f,n_{ef}}
\end{align}
depending on the orientation of the edge, relative to the face. The boundary amplitude \eqref{amplitudejn} can be written in the form of a path integral for an action
\begin{align}
S=\sum_f S_f=\sum_f \ln P_f
\label{action}
\end{align}
as
\begin{align}
W(j_l,n_{ne})=\sum_{\{j_f\}}\int dg_{ve}\int dn_{ef} e^{S}.
\label{amplitudejnS}
\end{align}
Notice also that the action can be always split in a bulk action (relative to internal faces) plus a boundary action (relative to external faces), so we can write it as
\begin{align}\label{amplitudejnSS}
W(j_l,n_{ne})=\sum_{\{j_f\}}\int dg_{ve}\int dn_{ef}\prod_l e^{i S_l}\prod_{f} e^{S_f}
\end{align}
where now the second product is over the sole internal faces.
\section{The scaling limit ($\hbar\rightarrow0$ regime)}
The continuum limit of the theory is defined as the infinite refinement of the 2-complex (possibly undergoing a second order phase transition), where we expect a large portion of spacetime to be described as the union of many small elementary regions (quanta) labeled by spins of order one. Differently, here we define a semiclassical regime that is suitable to describe gravitational physics with a truncation of the theory on a finite cellular structure of spacetime. The amplitude on a finite graph can be viewed as an effective amplitude associated to a coarse-graining procedure \cite{Markopoulou:2002ja,Bahr:2010cq} applied to the complete theory. We make the following hypothesis on the scaling limit of the spinfoam theory truncated to a fixed and finite 2-complex: the semiclassical limit is defined in the limit of uniform rescaling $\alpha$ of both boundary and bulk spin variables. The rationale behind this approximation is the following.

Notice that, restoring the $\hbar$ dependence in the theory, the relation between physical areas and spins is
\begin{align}
j\sim \gamma\frac{A}{\hbar}
\end{align}
so that in the semiclassical limit defined formally as the limit $\hbar\rightarrow 0$, all the spins become large. Observe also that in this limit the the minimal spin fluctuation allowed in the theory (of order one) becomes large. This is a consequence of the discreteness of the area spectrum.

The semiclassical limit will affect a quantum amplitude in the following way. Call $\Delta$ the spin spacing parameter ($\Delta$ is one before taking the semiclassical limit), and consider the boundary amplitude for a generic 2-complex 
\begin{align}
W_{\Delta}(j_l,n_{nl})
\end{align}
where we have inserted the explicit dependence on the spacing parameter. Define the semiclassical limit as the asymptotic regime
\begin{align}\label{scalinglimit}
W_{\alpha\Delta}(\alpha j_l,n_{nl})\quad\quad\alpha\rightarrow\infty
\end{align}
where $\alpha$ is a homogeneous rescaling parameter of the boundary spins and the spin step $\Delta$, or equivalently of both boundary and bulk spin quantum numbers. The correctness of this approximation has to be checked against concrete computations in specific examples, and possibly be justified and derived from the full amplitude (defined on the infinite 2-complex) as the result of the iteration of some kind of renormalization group transformation at the level of the spinfoam `lattice'. In this paper, we test this semiclassical regime for a simple fixed 2-complex.
\section{Large spins and geometry}
Here we review very briefly the asymptotic approximation \cite{Conrady:2008mk,Barrett:2009gg} of the amplitude at fixed and large spins, using the previous notations. Let us restrict our attention to a 2-complex which is dual to a simplicial triangulation. Now every vertex is bounded by five edges and ten faces, and every edge is bounded by four faces. Vertices are dual to 4-simplices, and edges are dual to tetrahedra.

Let us write \eqref{amplitudejnSS} as
\begin{align}
W(j_l,n_{ne})=\sum_{\{j_f\}} W(j_l,n_{ne};j_f).
\end{align}
The semiclassical analysis at fixed spins is the study of $W(j_l,n_{ne};j_f)$ when both $j_l$ and $j_f$ are large. More precisely, when they are uniformly rescaled with a large parameter $\alpha$. We can use an extended stationary phase method. The action $S$ is complex with real part $\text{Re}\,S\leq 0$, so the main contribution to the integral comes from critical points where $\text{Re}\,S=0$. It is easy to see that these are the solutions to
\begin{align}
g^+_{ev}\vec n_{ef}=-g^+_{e'v}\vec n_{e'f}\label{critical+}\\
g^-_{ev}\vec n_{ef}=-g^-_{e'v}\vec n_{e'f}
\label{critical-}
\end{align}
where $e,e'$ are adjacent edges in the face $f$, sharing the vertex $v$. If there are no solutions, the amplitude is exponentially suppressed. Using \eqref{critical+}, \eqref{critical-} we have that the brackets
\begin{align}
\bra{n_{ef}}(g_{ve}^{\pm})^{-1}g^\pm_{ve'}\varket{n_{e'f}}=e^{i\theta^\pm_{vf}}
\end{align}
reduce to simple phases, on the critical points (on-shell).

Furthermore, we must require that the critical points are also stationary. Varying $S$ with respect to $SO(4)$ group variables, and evaluating at a critical point we get the condition
\begin{align}
\delta_{g_{ev}} S|_{crit.}=0\;\longrightarrow\;\sum_{f\in e}j_{f}\vec n_{ef}=0
\end{align}
which expresses the closure relation for the tetrahedron dual to the edge $e$. Variation with respect to internal group elements $n_{ef}$ does not give further information, because it is automatically satisfied. 

The existence of critical points is related to the existence of a triangulation where areas of the triangles are specified by the set of spins $j_f$, and the unit directions normal to the triangles, in the 3-dimensional frames of tetrahedra, are specified by the set of unit vectors $n_{ef}$. Critical points have the interpretation of a 4-dimensional Regge manifold, that is a manifold endowed with a continuous, piecewise flat metric where curvature is distributional and concentrated on triangles. This can be seen as follows. Define the vectors
\begin{align}
\vec n^\pm_{vf}=g^\pm_{ev}\vec n_{ef}.
\end{align}
Then the vectors
\begin{align}
\vec J^\pm_{vf}=j_f^\pm\vec n^\pm_{vf}
\end{align}
can be interpreted as the selfdual ($+$) and antiselfdual ($-$) components of the discrete field
\begin{align}
J_{vf}=B_{vf}+\frac{1}{\gamma}{}^*B_{vf}\in so(4)
\end{align}
from which we can extract the field $B_{vf}$. The field $B_{vf}$ codes the spacetime metric degrees of freedom because
\begin{align}
A_{vf}=\frac{8\pi G\hbar}{c^3}\;{}^*B_{vf}
\end{align}
is the area bivector of the triangle $f$, in the frame of the 4-simplex $v$, where we have restored dimensional units. So $A_{vf}$ is the discretization of $e\wedge e$, where $e$ is the frame field tetrad. 

In the large spin limit, the action is evaluated at a critical point: $W(j_l,n_{nl};j_f)\sim\exp(S|_{crit.})$. The on-shell bulk action reads
\begin{align}
S_f|_{crit.}&=i\gamma j_f \Theta_f + ij_f \Theta_f^*
\label{Sfonshell}
\end{align}
where we have defined the angles
\begin{align}
&\Theta_f=\sum_{v\in f}(\theta^+_{vf}-\theta^-_{vf}),\\\label{Thetastar}
&\Theta^*_f=\sum_{v\in f}(\theta^+_{vf}+\theta^-_{vf}).
\end{align}
Remarkably, it turns out that $\theta_{vf}^+-\theta_{vf}^-$ is essentially the 4-dimensional dihedral angle between the two tetrahedra in the boundary of the 4-simplex $v$ that share the triangle dual to $f$, so that $\Theta_f$ is the deficit angle associated to the face $f$. Moreover, we have that $\Theta_{vf}^*$ vanishes. 

The geometric interpretation of \eqref{Sfonshell} is straightforward. The parallel transport $g_{ef}$ of the area bivector of a face $f$ from the reference frame attached to the 4-simplex $v=s(e)$ to the adjacent 4-simplex $v'=t(e)$ is defined by\footnote{Here $g\in SO(4)$ acts on the area bivector $A\in so(4)$ in the adjoint representation.}
\begin{align}
A_{v'f}=g_{ef} A_{ve}=g_{ev'}g^{-1}_{ev} A_{ve}.
\end{align}
This $SO(4)$ rotation (holonomy) splits into selfdual and antiselfdual rotations. The angle $\theta^*_{vf}=\theta_{vf}^++\theta_{vf}^-$ parametrizes a rotation in the plane of the triangle dual to $f$, while the dihedral angle $\theta_{vf}=\theta_{vf}^+-\theta_{vf}^-$ parametrizes a rotation in the plane \emph{orthogonal} to the same triangle, in the frame of the 4-simplex $v$. The first is a twisting angle: two tetrahedra that share a triangle can be `twisted' one with with respect to the other by a rotation along the direction normal to the triangle. The angle $\Theta_f^*$ \eqref{Thetastar} measures the total mismatch around a face, and can be thought as the discrete analog of torsion; since for large spins this angle is set to zero by the critical point equations, this means that the on-shell connection $g_{ef}$ is torsion-free, or in other words it is the discrete 4-dimentional spin connection. 

So the action \eqref{Sfonshell} is a discretization of the Holst action for gravity, when the on-shell connection is substituted. Indeed, recalling that the area spectrum of loop quantum gravity gives (for large spins) $A_f=\gamma j_f$, we have
\begin{align}
S^{bulk}|_{crit.}=i\sum_f A_f \Theta_f+\frac{1}{\gamma}i\sum_f A_f\Theta^*_f.
\end{align}
The first term is the Regge form \cite{Regge:1961px} of the action for discrete general relativity and the second term vanishes because the torsion angle is zero, in agreement with the fact that the second term vanishes on-shell also in the continuum theory.
\section{Many 4-simplices: a simple example}
\begin{figure}
\includegraphics[width=2.8cm]{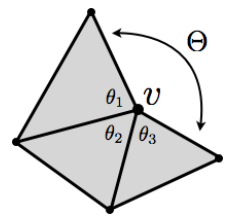}
\caption{Two-dimensional analog of $\sigma_3$. Curvature is concentrated at the vertex $v$ and is coded in the deficit angle $\Theta=2\pi-\theta_1-\theta_2-\theta_2$.}
\label{fig:2danalog}
\end{figure}

In this section we discuss the amplitude associated to a simple, but nontrivial triangulation in four dimensions. The main difference with respect to the analysis at fixed spins is that now we focus on the complete boundary amplitude $W(j_l,n_{nl})$, which contains the summation over internal spins. It will be also a fertile ground to test the viability of the scaling limit.
\begin{figure}
\vspace{0.3cm}
\includegraphics[width=3.5cm]{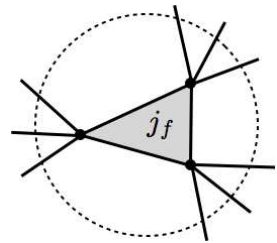}
\caption{Spinfoam diagram $\sigma_3$ dual to the simplicial triangulation obtained by glueing three 4-simplices on a common triangle. Two 4-simplices share a tetrahedron. The spinfoam face $f$ dual to the triangle is `colored' with the internal spin $j_f$.}
\label{fig:singleface}
\end{figure}
We consider a simplicial 2-complex $\sigma_3$ with a single internal face (see Fig.\ref{fig:singleface}), and the minimal number of internal edges, which is three. The complex $\sigma_3$ has three vertices. This corresponds to the most simple nontrivial triangulation of spacetime, obtained by glueing three 4-simplices pairwise through tetrahedra, so that they all share one triangle. This simple geometry is the 4-dimensional generalization of a 2-dimensional spacetime region (Fig.\ref{fig:2danalog}) triangulated with three triangles glued pairwise along one side, and curvature is concentrated at the common point.
This simple 4-dimensional  geometry allows us to test the dynamics of spinfoam models. Do curved spacetimes contribute to the boundary amplitude? In order to see this, let us describe in more detail the construction. The topology of the region is the one of a 4-ball and its boundary is a triangulation of the 3-sphere with nine tetrahedra (dual to the nodes of the graph). The boundary graph is drawn in Fig.\ref{fig:boundary}.

The boundary data are the set of 18 external spins $j_l$, one per each link in the boundary graph, and $4\times 9$ unit vectors $n_{nl}$, four per each of the nine nodes $n$. These data are chosen to be the boundary data of a Regge triangulation obtained by glueing three nondegenerate 4-simplices as described by the combinatorics of the spinfoam diagram of Fig.\ref{fig:singleface}. Classically, the 3-dimensional boundary geometry chosen determines uniquely the internal geometry. Curvature is concentrated at the internal triangle $f$ (the hinge) and is coded in the deficit angle $\Theta_f$.
\begin{figure}
\includegraphics[width=2.6cm]{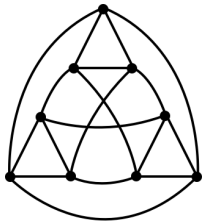}
\caption{Boundary graph of $\sigma_3$. It can be obtained by intersection of the 2-complex $\sigma_3$ with a topological 3-sphere, depicted as a dashed circle (see Fig.\ref{fig:singleface}).}
\label{fig:boundary}
\end{figure}

There is a naïve but worrying argument against the spinfoam state sum \eqref{amplitudejnSS}: given that the action \eqref{action} depends linearly on the spins, the terms proportional to the internal spins $j_f$ give rapidly oscillating phases which suppress the sum over $j_f$, unless their coefficients, which as we have seen have the interpretation of deficit angles $\Theta_f$, are zero. We can also say that the variation of the action with respect to the spin $j_f$ produces the equation $\Theta_f=0$. In other words, the spinfoam amplitude $W(j_l,n_{nl})$ seems to implement flatness in the bulk of the triangulation, as opposed to Ricci flatness, which is the correct Einstein equation in vacuum. As a remark, it has been shown in \cite{Mamone:2009pw} that the Barret-Crane spinfoam amplitude \cite{Barrett:1997gw} imposes flatness of the connection when semiclassical boundary wave-packets are considered. The main object of this paper is to see if this this problem persists in the `new models'.

However, there is one subtlety in the previous argument about the suppression of curved geometries:
\begin{itemize}
\item it demands stationarity of the action with respect to spin variables (see also \cite{Bonzom:2009hw}). But spins can take only discrete, integer values and we are not allowed to take derivatives with respect to them, unless we can show that this can be done in the limit we are considering.
\end{itemize}
We will show in a concrete example that this difficulty is only apparent, precisely because in the scaling limit the sum overs spins does not converge to a continuous integral, and therefore the equations of motion cannot be obtained by varying with respect to spins.

So let us study the boundary amplitude of $\sigma_3$ for large boundary spins $j_l$, and estimate the effective contribution of each term in the sum over internal spins in detail. For the simple case under consideration, the boundary amplitude \eqref{amplitudejnSS} has a single sum:
\begin{align}
W_{\Delta}(j_l,n_{nl})=\sum_{j_f\in\mbb N}\int dg_{ev}\int dn_{ef}e^{S_f}\prod_le^{S_l}.
\label{Wbulkboundary}
\end{align}
As before, we have split the action in bulk and boundary actions. We are interested in the semiclassical regime of the previous amplitude. We can apply the scaling limit \eqref{scalinglimit} defined previously. The amplitude \eqref{Wbulkboundary} in the large scale regime reads 
\begin{align}\nonumber
&W_{\alpha\Delta}(\alpha j_l,n_{nl})=\sum_{j_f\in \alpha \mbb N}W(\alpha j_l,n_{nl};j_f)=\\
&=\sum_{j_f\in\mbb N}W(\alpha j_l,n_{nl};\alpha j_f).
\label{largeW}
\end{align}
%
Following the asymptotic analysis at fixed spins \cite{Conrady:2008mk,Barrett:2009gg}, we conclude that the term $W_{\alpha\Delta}(\alpha j_l,n_{nl};\alpha j_f)$ is suppressed for $\alpha\rightarrow\infty$ unless $j_f$ takes a specific value $j_f=j^0_f$, because this is the only spin that can be associated to a Regge triangulation. Indeed consider one of the 4-simplices of the triangulation, or one of the spinfoam vertices. The boundary data (set of $j_l$'s and $n_{nl}$'s) fix nine areas of the 4-simplex, dual to the nine external faces bounding the vertex, and three sets of unit vectors associated to the three external edges. These data determine uniquely the the geometry of the 4-simplex (all of its ten lengths), and in particular the remaining internal area $j_f$, because we are overdetermining the geometry by fixing more than ten independent parameters. We can repeat the same argument for the other two vertices. This shows that there is a unique internal spin compatible with the boundary geometry. So when all spins are large, the contribution of the terms in \eqref{largeW} with $j_f\neq j^0_f$ is negligible.

Thus we find that for large spins,
\begin{align}
W(j_l,n_{nl};j_f)\sim\begin{cases}e^{i S_R} & j_f=j_f^0\\\text{suppressed} & j_f\neq j^0_f\end{cases}
\end{align}
with $S_R$ the Regge action, sum of a bulk action plus a boundary term:
\begin{align}
S_R=A_f\Theta_f+\sum_l A_l\Theta_l.
\end{align}
Here $A_l=\gamma j_l$, $A_f=\gamma j^0_f$ and $\Theta_f$ is the deficit angle at the internal face $f$. The boundary term contains the 4-dimensional dihedral angle $\Theta_l$ associated to the links, namely the angle between the 4-dimensional normals of the two tetrahedra attached by the triangle dual to the link $l$.

Now, a key observation is that the sum in \eqref{largeW} runs effectively over a finite set, due to the Clebsh-Gordan, or triangular inequalities associated to the internal edges. Indeed we have easily that the number of nonvanishing terms in the sum is bounded by $3 \alpha j_{max}$, where $j_{max}$ is the largest among the boundary spins $j_l$. This number grows linearly with $\alpha$. As a consequence, we can neglect a \emph{finite} number of exponentially suppressed terms, and keep only the term with $j_f=j_f^0$. So we have

\begin{align}
W_{\alpha\Delta}(\alpha j_l,n_{nl})=W_{\alpha\Delta}(\alpha j_l,n_{nl};\alpha j_f^0)+R(\alpha)\label{expreminder}
\end{align}
and the reminder is
\begin{align}
|R(\alpha)|\leq \alpha e^{-\alpha}.
\end{align}
Notice that the expansion \eqref{expreminder}, as we anticipated before, cannot be obtained from a stationary phase approximation of the sum over spins, because the summand does not approximate a continuous function in the large $\alpha$ limit. \footnote{Indeed the term $W(\alpha j_f,n_{nl};\alpha j_f^0+\alpha)$ with the internal spin closest to $\alpha j^0_f$ does not approach the dominant term $W(\alpha j_f,n_{nl};\alpha j_f^0)$. Actually, it decays exponentially.} In the large spin limit, obtained as the usual asymptotic expansion in $\hbar\rightarrow 0$, the spinfoam amplitude \eqref{Wbulkboundary} then reduces to
\begin{align}
W(j_l,n_{nl})\sim e^{i S_R}
\end{align}
and can be interpreted as the exponential of the Hamilton function for general relativity, namely the action evaluated on a solution of the equations of motion, viewed as a function of the boundary data. The nontrivial result we have found is that for a generic configuration of the boundary spins and boundary unit vectors (the boundary geometry) the path integral selects a spacetime triangulation with curvature, thus showing that the spinfoam model does \emph{not} impose flatness in the semiclassical limit, and possesses the correct local degrees of freedom.
\section*{Conclusions and outlook}
The simple example we presented here can be generalized to larger 2-complexes \cite{curvatureappear} and suggests that the spinfoam amplitude in the large scale limit takes the form of a sum
\begin{align}
W(j_l,n_{nl})\sim\sum_{\{j'_f\}} e^{i S_R}\label{generalization}
\end{align}
where the prime in $j'_f$ denotes a restriction on the sum to the spin configurations which arise as the areas of a Regge triangulation, and $S_R$ is the Regge action for those triangulations (it is well-known that a generic spin configuration does not correspond to any Regge triangulation \cite{Barrett:1997tx,Makela:2000ej,Dittrich:2008va}).

This is exactly what happens in the simple example studied in this paper: the internal spins $j_f$ which do not correspond to a Regge triangulation do not give contribution to the amplitude, and the effective amplitude takes the correct form \eqref{generalization}. The form \eqref{generalization} implies that in the semiclassical approximation the equations of motion are equivalent to those ones of Regge gravity \cite{Regge:1961px}, a discretization of Einstein equations. However, we must stress that our argument is more qualitative than quantitative, the main purpose being the identification of a possible mechanism for the emergence of classical general relativity from spinfoam quantum gravity, and we have disregarded other possible contributions to the amplitude (symmetry related spacetimes, vector geometries, etc.). For example we expect another term which corresponds to a spacetime with opposite orientation that would change the amplitude of $\sigma_3$ in
\begin{align}
W(j_l,n_{nl})\sim \cos(S_R)
\end{align}
as in the asymptotic formula for the amplitude of a single vertex \cite{PonzanoRegge:1968,Barrett:1998gs,Barrett:2009mw,Barrett:2009gg}. Finally, the amplitude of the three-vertex spinfoam $\sigma_3$ we have studied is well-defined (finite), but larger 2-complexes would give potentially divergent amplitudes in the `infrared'. Divergencies are associated to bubbles in the foam and a suitable regularization and renormalization scheme \cite{Perini:2008pd,Krajewski:2010yq,Geloun:2010vj,Rivasseau:2011xg} is required in order to recover a meaningful physics.
\section*{Acknowlegements}
This work was supported in part by the NSF grant PHY0854743, The George A. and Margaret M.~Downsbrough Endowment and the Eberly research funds of Penn State. E.M. gratefully acknowledges support from ``Fondazione Angelo della Riccia''.
\bibliography{bibliocurvature}
\end{document}